\documentclass[preprint,prd,groupedaddress,footinbib,showpacs,rotate]{revtex4}
\usepackage{amssymb,latexsym}
\usepackage{graphicx}
\usepackage{dcolumn}

\begin{document}

\title{ Deep inelastic structure functions from supergravity at small x}
\author{C.  A.  Ballon Bayona}
\email{ballon@if.ufrj.br} \affiliation{Instituto de F\'{\i}sica,
Universidade Federal do Rio de Janeiro, Caixa Postal 68528, RJ
21941-972 -- Brazil}
\author{Henrique Boschi-Filho }
\email{boschi@if.ufrj.br} \affiliation{Instituto de F\'{\i}sica,
Universidade Federal do Rio de Janeiro, Caixa Postal 68528, RJ
21941-972 -- Brazil}
\author{Nelson R. F. Braga}
\email{braga@if.ufrj.br} \affiliation{Instituto de F\'{\i}sica,
Universidade Federal do Rio de Janeiro, Caixa Postal 68528, RJ
21941-972 -- Brazil}


\begin{abstract}  
Deep inelastic structure functions can be calculated from 
supergravity when the Bjorken parameter $x$ satisfies 
$ x >  1/\sqrt{gN}\,$. 
We consider a gauge theory with very large 
't Hooft coupling $gN$  in order to investigate the region $x << 1$.
In this case the center of mass energy is large enough to increase the number of 
hadronic constituents of the final state. 
We calculate the structure functions in terms of the number of final hadronic constituents.      
At small $x$ we find a scaling law similar to geometric scaling but with  $\gamma_s = 0.5 $ 
and $ \lambda = 1 \,$. 
\end{abstract}


\maketitle

\section{ Introduction }

Deep inelastic lepton-hadron scattering (DIS) is  an important experimental tool for 
investigating the structure of hadrons. The cross sections for this process
can be represented in terms of structure functions depending on the kinematical 
scalar variables $q^2$ and $x$ where $q^2 $  is the virtuality of
the photon emitted by the charged lepton and $x $ is the Bjorken variable. 
For large $x$ the structure functions are approximately independent of $q^2$. This is known as Bjorken
scaling. For scattering at small $x$ ( $x < 0.01 $) the observed 
total cross section depends on $x$ and $q^2$ through the variable: 

\begin{equation}
{\cal T} = q^2 x^\lambda / q^2_0 \, x_0^\lambda \,
\end{equation}

\noindent  where $ q_0 = 1 GeV $ and $x_0 = 3 \times 10^{-4}$ and
 $ 0.3 < \lambda < 0.4 $ \cite{Stasto:2000er}. This is called geometric scaling.
A study of this phenomena from the point of view of perturbative QCD can be found, for instance, in 
\cite{Iancu:2002tr,Munier:2003vc,Iancu:2003ge}.

A description of DIS at large coupling from the point of view of 
gauge/string duality was formulated by Polchinski and Strassler
in \cite{Polchinski:2002jw}. They calculated the DIS structure functions considering a 10-d 
$AdS_5\times W$  space with a hard infrared cut off that breaks conformal invariance. 
Three different kinematical regimes were considered: 

\noindent $\bullet$ $ x >  1/\sqrt{gN}\,$ 

\noindent $\bullet$ $ \,\exp \,( - \sqrt{gN }) << x << (gN)^{-1/2}\,\,$

\noindent $\bullet$ $\, x <  \exp \,( - \sqrt{gN }) \,$ .

\noindent In the first regime the supergravity approximation is valid while in the other two regimes 
massive string states contribute.

The introduction of a hard cut off in AdS space leads to a discrete mass spectrum 
for hadrons \cite{Boschi-Filho:2002vd,deTeramond:2005su,Erlich:2005qh} with asymptotically quadratic
Regge trajectories.
An alternative approach to introduce an infrared cut off, that leads to linear Regge trajectories for
the hadronic  spectrum is the soft wall model \cite{Karch:2006pv,Colangelo:2007pt}. The DIS structure functions were calculated recently using this 
model \cite{BallonBayona:2007qr}.
Hadronic form factors were also calculated using gauge string duality in for example  
\cite{Brodsky:2007hb,Grigoryan:2007vg,Grigoryan:2007my}.  

The DIS structure functions obtained in ref. \cite{Polchinski:2002jw} 
(and also in\cite{BallonBayona:2007qr}), from supergravity, were calculated considering that the final hadronic 
operator has a conformal dimension $\Delta'$ equal to the conformal dimension of the initial hadronic 
operator $\Delta$. 
As seen in \cite{Polchinski:2001tt}, the conformal dimension of a hadronic operator can be identified with the minimum number of constituents of the hadronic state. So the case $\Delta' = \Delta\,$ represents scattering processes where the final state has the same number of constituents of the initial state. 

The initial state in a DIS process contains one hadron. 
When the center of mass energy $ \sqrt{s}\,$ is larger than the hadronic mass scale
we expect more hadrons in the final state. 
Since $ s \sim q^2 / x\,$, this corresponds to small  $x$. 
The increase in the number of  hadrons should be associated with an increase in the number of constituents
of the final state.  

In the standard AdS/CFT dictionary one associates a bulk supergravity field for each boundary state.
So we should introduce one additional bulk field  for each new hadron in the final state.
We will follow here a simpler phenomenological approach that consists in using just one supergravity
field for the final state. The boundary operator dual to this field will have a conformal dimension
$\Delta' > \Delta $, representing a final state with more constituents than the initial state.
We conjecture that this state evolves to multi-hadronic states.

We calculate within the supergravity approximation the contributions to the DIS structure 
functions from final states with $\Delta' > \Delta $. 
We show that these contributions are relevant in the small $x$ regime
where they change the structure function dependence on the variables $q^2 $ and $x$
with respect to the results obtained in refs. \cite{Polchinski:2002jw,BallonBayona:2007qr} where
$\Delta' = \Delta\,$. 
We find a scaling law for the total cross section  similar to geometric scaling.


\section{ Structure functions from supergravity  }

In deep inelastic scattering a lepton produces a virtual photon of 
momentum $q^\mu$ which interacts with a hadron of momentum $P^\mu $ (see Fig. 1). 
The experiment detects the final lepton, determining the momentum transfer $q^\mu$, but not the final hadronic state $X$, just its four-momentum $P_X^\mu$.      
We can parameterize the process using as dynamical variables the virtuality $q^2$ and the Bjorken parameter $x \equiv  -q^2 /2P\cdot q \,$. 

\begin{figure}
\setlength{\unitlength}{0.1in}
\vskip 3.cm
\begin{picture}(0,0)(10,0)
\rm
\thicklines
\put(1,14.5){$\ell$}
\put(3,15){\line(2,-1){7}}
\put(3,15){\vector(2,-1){4}}
\put(18,14.5){$\ell$}
\put(17,15){\line(-2,-1){7}}
\put(10,11.5){\vector(2,1){4.2}}
\put(9.5,8){$q$}
\bezier{300}(10,11.5)(10.2,10.7)(11,10.5)
\bezier{300}(11,10.5)(11.8,10.3)(12,9.5)
\bezier{300}(12,9.5)(12.2,8.7)(13,8.5)
\bezier{300}(13,8.5)(13.8,8.3)(14,7.5)
\put(0,-2){$P$}
\put(3,0){\line(2,1){10.5}}
\put(3,0){\vector(2,1){6}}
\put(16,6){\circle{5}}
\put(27,-2){$X$}
\put(18.5,5.5){\line(3,-1){8}}
\put(18.3,5){\line(2,-1){8}}
\put(18,4.5){\line(3,-2){7.5}}
\put(17.5,3.8){\line(1,-1){6}}
\end{picture}
\vskip 1.cm
\parbox{4.1 in}{\caption{} Illustrative diagram for a deep inelastic scattering. A lepton $\ell$ exchanges a virtual photon with a hadron of momentum $P$.}
\end{figure}
\vskip .5cm

The deep inelastic hadronic tensor for unpolarized scattering is defined as 
\begin{equation}
W^{\mu\nu} \, = i \, \int d^4y\, e^{iq\cdot y} \langle P, {\cal Q} \vert \, \Big[ J^\mu (y) , J^\nu (0) \Big] 
\, \vert P, {\cal Q} \rangle \,,
 \label{HadronicTensor}
\end{equation}

\noindent where $ J^\mu(y)$ is the electromagnetic hadron current and $ {\cal Q} $ is the electric charge of the initial  hadron. This tensor can be decomposed into the  structure functions $F_1 (x,q^2) $ and $F_2 (x,q^2) $ as \cite{Manohar:1992tz}
\begin{equation}
W^{\mu\nu} \, = \, F_1 (x,q^2)  \Big( \eta^{\mu\nu} \,-\, \frac{q^\mu q^\nu}{q^2} \, \Big) 
\,+\,\frac{2x}{q^2} F_2 (x,q^2)  \Big( P^\mu \,+ \, \frac{q^\mu}{2x} \, \Big) 
\Big( P^\nu \,+ \, \frac{q^\nu}{2x} \, \Big)
\, ,  \label{Structure}
\end{equation}

\noindent where $\eta_{\mu\nu}={\rm diag}(-,+,+,+)$. 

The DIS cross section is related to the forward Compton scattering tensor  
\begin{equation}
T^{\mu\nu} \, = i \, \int d^4y e^{iq\cdot y} \langle P, {\cal Q} \vert \,  T \Big(  J^\mu (y) J^\nu (0) \Big)  
\, \vert P, {\cal Q} \rangle\,,
 \label{forwardamplitude}
\end{equation}

\noindent which can be decomposed as 
\begin{equation}
T^{\mu\nu} \, = \, {\tilde F}_1 (x,q^2)  
\Big( \eta^{\mu\nu} 
\,-\, \frac{q^\mu q^\nu}{q^2} \, \Big) 
\,+ \,\frac{2x}{q^2} {\tilde F}_2 (x,q^2)  
\Big( P^\mu \,+ \, \frac{q^\mu}{2x} \, \Big) 
\Big( P^\nu \,+ \, \frac{q^\nu}{2x} \, \Big)
\, ,  \label{CompStructure}
\end{equation}

\noindent where ${\tilde F}_1 (x,q^2)  $ and ${\tilde F}_2 (x,q^2) $ are the associated structure functions. 

The optical theorem relates the tensors $W^{\mu\nu}$ and $T^{\mu\nu}$ and implies
\begin{equation}
\label{optical}
F_{1,2} (x,q^2) \equiv 2 \pi \,{\rm Im }\,{\tilde F}_{1,2} (x,q^2)\,.
\end{equation}

The imaginary part of $T^{\mu\nu}\,$ can be expressed in terms of a sum over the intermediate states $X$ 
formed in the hadron-photon collision
\begin{equation}
\label{Imag}
{\rm Im} T^{\mu\nu} \, = \,  2 \pi^2 \, \sum_X \,\delta \Big( (P+q)^2 \,- P_X^2 \, \Big) 
\langle P, {\cal Q} \vert J^\nu ( 0 )   \vert P + q,\, X \rangle\,
\langle P + q , \, X \vert J^\mu ( 0 )   \vert P, {\cal Q} \rangle\,\,.
\end{equation}

Polchinski and Strassler\cite{Polchinski:2002jw}
found prescriptions for calculating this quantity from gauge string duality.
When the Bjorken variable satisfies the condition $ x > (gN)^{-1/2}\,$ one can use a supergravity approximation for 
string theory, since only massless string states contribute.  In this work we will be restricted to 
this regime. The prescription for the case of the scattering of a scalar particle by a virtual photon of polarization 
$\eta_\mu\, $ takes the form  
\begin{equation}
(2 \pi)^4 \, \delta^4 ( P_X - P - q ) \,\eta_\mu \,  \langle P + q, X \vert J^\mu ( 0 )  \vert P,{\cal Q} \rangle\,=\,  \int d^{10}x \sqrt{-g} A^m\,j_m \,,
\label{INTERACTION}
\end{equation}

\noindent where 
\begin{equation}
j_m \,=\, i  \, {\cal Q} \Big( \Phi_i\partial_m \Phi_X^\ast \,-\, 
  \Phi_X^\ast  \partial_m \Phi_i     \Big)\,
\label{5Current}
\end{equation}

\noindent is a five dimensional scalar current and  $A_m =(A_z,A_\mu)$ is a Kaluza-Klein gauge field. $ \Phi_i  $ and $ \Phi_X$ are dilaton fields representing the initial and final scalar states. 
All these fields live in the 10-d
 space $AdS_5 \times W \,$, 
with metric $g_{MN}$: 
\begin{equation}
\label{AdS} ds^2 \equiv g_{MN} \,dx^M dx^N \,= \, \frac{R^2}{z^2}( dz^2  +
\eta_{\mu\nu} dy^\mu dy^\nu  )\,+  \,R^2 ds_W^2\,\,\,, 
\end{equation}

\noindent where $W $ is a compact space. The hard wall condition consists of restricting the radial coordinate
$z $ to the region: $ 0 < z < 1/\Lambda \,$, where $\Lambda $ is the minimum hadronic mass in 
the four dimensional gauge theory. 

The equations of motion for the gauge field in the $AdS_5 $ space are
\begin{eqnarray}
\Box A_z &-& \partial_z \partial_\mu \Big( \eta^{\mu\nu}  A_\nu  \Big) \,=\, 0
\label{Eqmotion1}
\\
z \partial_z \Big( \frac{1}{z} \partial_z (\eta^{\mu\nu}  A_\nu ) \Big) 
&+& \Box ( \eta^{\mu\nu}  A_\nu  ) \,-\, \eta^{\mu\nu} \partial_\nu 
\Big( z  \partial_z ( \frac{1}{z} A_z )
\,+\, \partial_\alpha ( \eta^{\alpha\beta}  A_\beta  ) \Big) \,=\, 0
\label{Eqmotion2}
\end{eqnarray}

In order to represent a photon with momentum $q^\mu $ and polarization
$\eta^\mu $ on the boundary one imposes 
\begin{equation}
A_\mu (z=0, y) = \eta_\mu e^{i q\cdot y }\,.
\end{equation}

\noindent The polarization can be decomposed into longitudinal and transversal components:
$ \eta_\mu =  \eta_\mu^{^L } +  \eta_\mu^{^T}  $ with
\begin{equation}
\eta_\mu^{^L} \,=\, \frac{( \eta \cdot q ) }{q^2} q_\mu \,\,\,\,,\,\,\,\,
\eta_\mu^{^T} \,=\, \eta_\mu \,-\,  \frac{(\eta \cdot q ) }{q^2} q_\mu \,.
\end{equation}  

A natural ansatz for solving the equations of motion with this boundary condition is 
\begin{equation}
A_\mu (z , y) = \Big( \eta_\mu^{^L } f_{_L} (z) \, +\, \eta_\mu^{^T} f_{_T} (z) \Big)
e^{i q\cdot y }\,,
\end{equation}
 
\noindent where $f_{_L} (z)$ and $f_{_T} (z) $ are functions satisfying the boundary condition $ f_{_T} (0) =  
f_{_L} (0)= 1 $. 

The equations of motion (\ref{Eqmotion1}) and (\ref{Eqmotion2}) show us that 
$A_z $ has the same plane wave dependence as $\,A_\mu \,$.  The first equation gives
\begin{equation}
 A_z \,=\, - \frac{ i}{ q^2}  \, ( q \cdot \eta^{^L} )  \, e^{i q\cdot y } \,\partial_z f_{_L}(z)\,
=\, - \frac{ i}{ q^2}  \, q^\mu \partial_z A_\mu \,\,.
\label{e3}
\end{equation}  

Inserting this solution in eq. (\ref{Eqmotion2}) we find
\begin{equation}
\eta^{^T }_{\mu } 
\,  \Big[ z \partial_z \Big( \frac{1}{z} \partial_z  f_{_T}(z) \Big) 
\,-\,q^2 f_{_T}(z) \Big]\,=\,0
\label{e4}
\end{equation}  

\noindent with solution $\, f_{_T}(z) \,=\, q z  \, K_1(qz) \,$, where $ K_1 $ is a modified Bessel function. 

There is no equation for $ f_{_L}(z) $. So, the longitudinal part of the gauge field $A_\mu $ and the radial component
$A_z$ remain undetermined. These components can be determined by imposing a gauge condition.
Polchinski and Strassler used the convenient gauge condition
\begin{equation}
\label{Gaugecondition}
\eta^{\mu\nu} \partial_\mu A_\nu  \,+\,
z \partial_z \Big(  \frac{1}{z}
 A_z \Big) \,=\,0 \,,
\end{equation}
 
\noindent that leads to $ f_{_T}(z)\,=\, f_{_L}(z)\,\,  $ and implies the solutions
\begin{eqnarray}
A_{\mu}& = & \eta_{\mu}e^{iq\cdot y}\, qz K_1(qz) \nonumber \\
A_z & = & - i q \cdot \eta \, e^{iq\cdot y} \, z \, K_0(qz)\,.
\end{eqnarray}

Another possible gauge choice would be $ \partial_\mu A^\mu \,= 0$ that corresponds to a transverse photon
($ \eta^{^L}_{\mu} \,=\, 0$) and implies $A_z = 0 $.

For the dilaton fields, the $AdS_5$ part of the equation of motion reads

\begin{equation}
\frac{1}{ \sqrt{ -g }} \, \partial_m \Big( \sqrt{ -g } \,\partial^m \Phi \,\Big) \,-\,m_5^2 \Phi
 =\,0 \,.
\end{equation}
 
\noindent where the five dimensional mass is related to the conformal dimension $\Delta $  of the boundary operator dual to $\Phi$ by

\begin{equation}
m_5^2 \,=\, \frac{\Delta (\Delta - 4) }{ R^2 } \,.
\end{equation}
 
The solutions for Dirichlet boundary conditions at $ z = 1/ \Lambda\,$ for the initial and final scalar states 
with four dimensional momenta $P^\mu $ and $P^\mu_X$ are given by
 
\begin{eqnarray}
\Phi_i &= &  e^{iP\cdot y}\frac{C_i}{R^4}  \Lambda \, z^{2}\, J_{\Delta - 2}( \Lambda z)  Y(\Omega) \nonumber \\
\Phi_X &= &  e^{iP_X\cdot y} \frac{C_X }{R^4} \, \Lambda^{1/2}s^{1/4}z^2 \, J_{\Delta'-2}(s^{1/2}z)Y(\Omega) 
\,, \label{scalarfielddelta}
\end{eqnarray}

\noindent where $ s = - P_X^2 \,, P^2 = - \Lambda^2 \,,$  $C_i $ and $C_X $ are normalization constants and 
$Y(\Omega) $ is the angular part of the solution, that is an eigenstate of the Laplacian in the space $W$. 

The quantities $\Delta $ and $\Delta' $ are the conformal dimensions of the operators that create the 
initial and final boundary states. These dimensions can be identified with the number of constituents of the states\cite{Polchinski:2001tt}.
Polchinski and Strassler considered the case $\Delta' \,=\, \Delta $ that can be  
interpreted as a DIS process where the number of constituents is preserved. 

The condition $\Delta' \,=\, \Delta $ implies the conservation of the five dimensional current 
 $ j_m = ( j_z , j_\mu )$ defined in eq. (\ref{5Current}):
\begin{equation}
\frac{1}{ \sqrt{ -g }} \, \partial_m \Big( \sqrt{ -g } \,j^m \,\Big)  =\,0 \,.
\end{equation}

Note that the four dimensional part of this current is not conserved: $ \partial_\mu j^\mu \, \ne 0\,$.
Nevertheless, the supergravity interaction between $A^m $ and 
$j_m $ is  equivalent to an interaction between the 4-d $A_\mu $ and an effective 4-d conserved current:
\begin{equation}
\label{jeff}
 \int d^{10}x \sqrt{-g} A^m\,j_m \,= \int d^{10}x \sqrt{-g} A^\mu\,
\Big( j_\mu \,-\, \frac{ i}{ q^2}  q_\mu  \eta^{\nu\alpha} \partial_\nu j_\alpha   \Big) \,
\,\equiv  \int d^{10}x \sqrt{-g} A^\mu\,j^{eff}_\mu \,,
\end{equation}

\noindent with $\,\,\partial_\mu j_{eff}^\mu \,=\,0\,\, $ (using $q_\mu = (P_X - P)_\mu \,$). 
This conservation leads to a 
$ T^{\mu\nu}$ that satisfies the transversality conditions $  q_\mu T^{\mu\nu} = 0 = q_\nu T^{\mu\nu}$
as in eq. (\ref{CompStructure}). 

The structure functions obtained in ref. \cite{Polchinski:2002jw} are 

\begin{equation}
\label{FEPS}
F^{^{\Delta' = \Delta}}_1 (x,q^2)\,=\,0 \,\,\,;\,\,\, F^{^{\Delta' = \Delta}}_2 (x,q^2)\,=\, 
\pi C_0 \, {\cal Q}^2 \left( \frac{\Lambda^2}{q^2} 
\right)^{\Delta - 1}  x^{\Delta + 1} (1- x)^{\Delta - 2}\,.
\end{equation}

\noindent where $C_0 = 2^{2\Delta} \pi \vert C_i \vert^2 \vert C_X \vert^2 \, \Gamma^2 (\Delta )\,$.

It is interesting to note that the above structure function presents Bjorken scaling for $\Delta =1$, which means that, in this case, the hadron would have only one constituent. Experimentally this behavior is observed for $x > 0.1$.

On the other side, the above structure function at small $x$ have the following behavior 
\begin{equation}
\label{PS1}
\frac{ F^{^{\Delta' = \Delta}}_2 (x,q^2)}{q^2} \,\sim \, \Big( q^2\, x^{ \lambda }  \,
\Big)^{\gamma_s - 1 }\,, 
\end{equation}

\noindent with $ \lambda = - \frac{\Delta+ 1}{\Delta}\,$ and $\gamma_s = 1 -  \Delta $.

\section{ The $ \Delta' > \Delta $ case}

In a deep inelastic experiment at small $x$ the center of mass energy squared  $ s \approx q^2 / x\,$ is large enough to produce multi-hadronic final states. 
A complete description of such processes should involve introducing more bulk fields to represent these final states.
We will not attempt to do this here. We will consider a "final state"  still represented by just one dilaton field $ \Phi_X$, but with more constituents than the initial state. That means: $ \Delta' > \Delta $.
Eventually this state will evolve to produce multiple hadrons. 

When $ \Delta' > \Delta \,$ the 5-d current defined by eq (\ref{5Current}) is not conserved :

\begin{equation}
\frac{1}{ \sqrt{ -g }} \, \partial_m \Big( \sqrt{ -g } \,j^m \,\Big)  = \frac{ i \cal Q }{R^2}\Phi_i\Phi^{\ast}_X [\Delta'(\Delta'-4)-\Delta(\Delta-4)] \ne 0\,\,.
\end{equation}

If we use this non-conserved current to define an interaction of the type $A^m j_m $ as in the previous section,
we would  find an effective four dimensional current $j^{eff}_\mu $ that is non-conserved too.
So, we would end up with a non transverse $  T^{\mu\nu} $\footnote{In spite of this non transversality, it would be  possible to extract the structure function $F_2 $ from  $  T^{\mu\nu} $ by choosing a transverse photon.  This would correspond to choosing the  gauge condition $ \partial_\mu A^\mu \,=\,0$.}.

We will follow a phenomenological approach and define a modified five dimensional hadronic current:
\begin{equation}
\tilde{j}_m \, \equiv \, j_m - \, v_m \,
\frac{1}{ \sqrt{ -g }} \, \partial_n \Big( \sqrt{ -g } \,j^n \,\Big) 
\,.
\end{equation}

\noindent A possible choice of $v_m$ that implies a conservation of  $ \tilde{j}_m $ is
\begin{equation}
v_z \,=\, 0 \,\,\,\,\,\,;\,\,\,\,\,\,\,  
v_\mu \,=\,  i \, \frac{(P_X-P)_{\mu}}{(P_X-P)^2} \frac{R^2}{z^2}\,\,.
\end{equation}

\noindent This current leads to a  transverse $  T^{\mu\nu} $. 
Our prescription for the four dimensional current matrix element is 
\begin{eqnarray}
& & (2 \pi)^4 \, \delta^4 ( P_X - P - q )  \, \eta_\mu \,  \langle P + q, X \vert J^\mu ( 0 )  \vert P,{\cal Q} 
 \rangle\,\,=\, \int d^{10}x \sqrt{-g} A^m\,\tilde{j}_m \,. 
\label{NewInteraction}
\end{eqnarray}

\noindent This interaction term reduces, as in the previous section, to an interaction with an effective
 four dimensional conserved current
\begin{eqnarray}
& & \int d^{10}x \sqrt{-g} A^m\,\tilde{j}_m \,= \int d^{10}x \sqrt{-g} A^\mu\,j^{eff}_\mu \,
\nonumber\\
& & =\,  (2 \pi)^4 \, \delta^4 ( P_X - P - q ) \, 2{ \cal Q} C_iC_X \, s^{1/4}\Lambda^{\Delta-1/2} \, q \, \eta_{\nu}\big(p^{\mu}+\frac{q^{\mu}}{2x} \big ) {\cal I}\,,
\label{NewInteraction2}
\end{eqnarray}

\noindent where $\,j^{eff}_\mu \,$ is again defined by eq. (\ref{jeff}) and  
\begin{equation}
{\cal I } \, \approx  \,  
\int_0^{1/\Lambda} dz \, z^{\Delta} J_{\Delta'-2}(s^{1/2}z)K_1(qz) 
\approx \int_0^{\infty} dz \,z^{\Delta} J_{\Delta'-2}(s^{1/2}z)K_1(qz)\,\,.
\end{equation}

The result for this integral is 
\begin{eqnarray}
{\cal I} & = & 2^{\Delta-1}\frac{\Gamma(\frac{\Delta'+\Delta}{2})\Gamma(\frac{\Delta'+\Delta}{2}-1)}{\Gamma(\Delta'-1)}\big(\frac{s}{q^2}\big)^{\Delta'/2-1}(q^2)^{-(\Delta+1)/2} \, \nonumber \\& \times&F(\frac{\Delta'+\Delta}{2} , \frac{\Delta'+\Delta}{2}-1 ; \Delta'-1 ; -\frac{s}{q^2}) 
\end{eqnarray}

\noindent where $F(a,b\,;c\,;\omega)$ is the Gauss hypergeometric function defined by 
\begin{equation}
F(a,b\,;c\,;\omega) = \sum_{n=0}^{\infty} \, \frac{(a)_n \, (b)_n}{(c)_n} \, \frac{x^n}{n!} \, ,
\label{hypseries}
\end{equation}
 
\noindent with $(a)_n = \Gamma(a+n)/\Gamma(a)$.

Using these results in eq. (\ref{Imag})  we find the contributions to the structure functions
coming from final states with dimension $\Delta'$: 
\begin{eqnarray}
F^{^{\Delta'}}_1(x,q^2) &=&  0\nonumber\\
F^{^{\Delta'}}_2(x,q^2) & = & \pi^2 \, 2^{2\Delta}|C_i|^2|C_X|^2{\cal Q}^2 \big(\frac{\Lambda^2}{q^2}\big)^{\Delta-1}x^{1-\Delta'}(1-x)^{\Delta'-2} \, \big[ \frac{\Gamma(\frac{\Delta'+\Delta}{2})\Gamma(\frac{\Delta'+\Delta}{2}-1)}{\Gamma(\Delta'-1)} \big ]^2\, \nonumber \\ &\times&\big [F(\frac{\Delta'+\Delta}{2} , \frac{\Delta'+\Delta}{2}-1 ; \Delta'-1 ; -\frac{1-x}{x}) \big ]^2 \,, 
\end{eqnarray}

\noindent  where we have used the hard wall result for the sum of delta functions 
\begin{equation}
\label{nova3}
\sum_X \,\delta \Big( (P+q)^2 \,-\, P_X^2   \Big) 
\,=\, \frac{1}{2\pi s^{1/2} \,\Lambda\,}\,.
\end{equation}

Note that the structure functions will involve the sum over all possible values of $\Delta'\,$
\begin{equation}
F_2(x,q^2)\,=\,  F^{^{\Delta' = \Delta}}_2(x,q^2) \,+\, \sum_{\Delta' > \Delta } F^{^{\Delta' }}_2(x,q^2)  \,.
\end{equation}

Now we introduce the variable $\rho\equiv (\Delta'-\Delta)/2$ to represent the number of extra hadrons (mesons) that can be produced by the "final state".  
The minimum number of constituents of a hadron is 2. So, we expect non-negative integer numbers for $\rho$.  For example ,  for $\rho=0$  the final state can not produce extra hadrons, while for $\rho=1$ we can expect the production of an extra meson by the "final state".  The structure function expressed in terms of $\rho$ is
\begin{eqnarray}
F^{\,\rho\,}_2(x,q^2) & = & \pi^2 \, 2^{2\Delta}|C_i|^2|C_X|^2{\cal Q}^2 \big(\frac{\Lambda^2}{q^2}\big)^{\Delta-1}x^{1-\Delta}(1-x)^{\Delta-2} \, \big(\frac{1-x}{x}\big )^{2\rho}\big[ \frac{\Gamma(\Delta+\rho)\Gamma(\Delta+\rho-1)}{\Gamma(\Delta+2\rho-1)} \big ]^2\, \nonumber \\
& &\times \big[F(\Delta+\rho \, , \Delta+\rho-1 \, ; \Delta+2\rho-1 \, ; \, -\frac{1-x}{x}) \big ]^2 \,.
\end{eqnarray}

Using the property  
\begin{equation}
F(a,b;c;\omega) \, = \, (1-\omega)^{-a} \, F(a,c-b;c;\frac{\omega}{\omega-1})\,, 
\end{equation}

\noindent we have 
\begin{equation}
F^{\,\rho\,}_2(x,q^2) \, = \, F_2^{(\rho=0)}(x,q^2)\big [\frac{\Gamma(\Delta+\rho)\Gamma(\Delta +\rho-1)}{\Gamma(\Delta)\Gamma(\Delta +2\rho-1)} \big]^2 \, (1-x)^{2\rho} \, \big[F(\Delta+\rho \, , \rho \, ; \Delta+2\rho-1 \, ; \,  1-x) \big ]^2 \,,\label{rhostructurefunctions}
\end{equation}

\noindent where $F_2^{(\rho=0)}(x,q^2)$ is the structure function of eq. (\ref{FEPS}) obtained in
\cite{Polchinski:2002jw}. 
 
The $\rho=1$ case (one extra meson) is simple to calculate because the first and third arguments of the hypergeometric function are equal and simplify in the hypergeometric series of eq. (\ref{hypseries}) . We find  

\begin{equation}
F_2^{(\rho=1)}(x,q^2)\,  = \, F_2^{(\rho=0)}(x,q^2) \, x^{-2}(1-x)^2 \, . \label{stfuncrho1}
\end{equation}

The $\rho=2$ calculation is more subtle. From (\ref{rhostructurefunctions}) we have that 

\begin{eqnarray}
F_2^{(\rho=2)}(x,q^2) &=& F_2^{(\rho=0)}(x,q^2)\big(\frac{\Delta}{\Delta+2}\big)^2 \, (1-x)^4 \, \big[ \sum_{n=0}^{\infty} \frac{(\Delta+2)_n \, (2)_n}{(\Delta+3)_n} \, \frac{(1-x)^n}{n!} \big]^2 \nonumber \\
&=& F_2^{(\rho=0)}(x,q^2)\, \Delta^2(1-x)^4 \, \big[ \sum_{n=0}^{\infty} \frac{(2)_n}{n!} \, \frac{y^n}{(\Delta+2+n)} \big]^2\label{structurefunctionrho2}
\end{eqnarray}
 
\noindent with $y\equiv 1-x $ . The above series can be re-expressed as the integral:
\begin{eqnarray}
\sum_{n=0}^{\infty} \frac{(2)_n}{n!} \, \frac{y^n}{(\Delta+2+n)} &=& \frac{1}{y^{\Delta+2}}\sum_{n=0}^{\infty} \frac{(2)_n}{n!} \, \int_0^{y}dy' \, (y')^{\Delta+1+n} \nonumber \\
&=&\frac{1}{y^{\Delta+2}}\int_0^{y}dy' \, (y')^{\Delta+1}\,  \big[\sum_{n=0}^{\infty} \frac{(2)_n}{n!} (y')^n \big] \, \nonumber \\
&=&\frac{1}{y^{\Delta+2}} \, \int_0^{y}dy' \, (y')^{\Delta+1} \, (1-y')^{-2} \,. \label{seriesrho2}
\end{eqnarray}

Defining now $x' \equiv 1-y'$ and substituting (\ref{seriesrho2}) in (\ref{structurefunctionrho2}) we obtain 

\begin{eqnarray}
F_2^{(\rho=2)}(x,q^2) &=&  F_2^{(\rho=0)}(x,q^2) \, \Delta^2 \, (1-x)^{-2\Delta} \big[\int_x^{1} \, dx' \, (x')^{-2} \, (1-x')^{\Delta+1} \big]^2\, \nonumber \\
&=& F_2^{(\rho=0)}(x,q^2) \, \Delta^2 \, (1-x)^{-2\Delta} \, {\cal S}^{\,2} \label{stfuncrho2}\,
\end{eqnarray}

\noindent where 
\begin{equation}
{\cal S} \equiv -1 + \frac{1}{x} +(\Delta+1)\ln{x} + \sum_{n=2}^{\Delta+1}\pmatrix{\Delta+1\cr n\cr } \frac{(-1)^n}{(n-1)}\big[1-x^{n+1}\big] \, ,
\end{equation}

\noindent and 
\begin{equation}
\pmatrix{\Delta+1\cr n } = \frac{(\Delta+1)!}{n! \, (\Delta+1-n)!}\,.
\end{equation}

The cases $ \rho \ge 3 $ can be calculated in a similar way. 
So it is possible to calculate the contributions for final states corresponding 
to different numbers of extra hadrons. It is interesting to observe what happens when $x\to 1$ in (\ref{stfuncrho1}) and (\ref{stfuncrho2}).
In this limit all the contributions to $F_2 $ vanish, including the case $\rho = 0 $. 
Defining $ \epsilon = 1 - x \,\to 0 \,$ it is not difficult to show that 
the structure functions contributions from $\rho = 1 , 2 $ involve powers of $\epsilon $ that are higher than the $\rho = 0$ contribution. 
So they are negligible near this limit and the structure function can be approximated
by considering just $\Delta' = \Delta$. This is consistent with the fact that the case $ x= 1 $ is the elastic scattering limit where no extra hadrons are produced by the final state.
In the next section we will investigate the small $x$ region, where the contribution of final states that 
produce extra hadrons will be important.


\section{Small ${\boldmath x}$ and multi-hadronic states}

The supergravity approximation is valid if $ x > (gN)^{-1/2}\,$  so if the 't Hooft constant $gN$
is large enough, we can still use this approximation for a region where $ x << 1$.
In this regime the center of mass energy squared $s \approx q^2 /x $ is very large. 
So we expect the production of a large number of hadrons.

In our formulation there is only one field $\Phi_X $ that represents the product of the scattering process.   We conjecture that this state, that has a large number of constituents, will evolve into multi-hadronic states, but we will not describe this evolution. This would require the inclusion of more fields in the interaction action.

For a fixed value of the energy and momentum of the final state, the maximum number of hadrons that can be produced by the final state occur when there is no relative motion and all the extra hadrons have the minimum mass $\Lambda$.
Taking the approximation that the initial hadron has also a mass $\Lambda$, the maximum number of produced
hadrons is

\begin{equation}
\label{Nmax}
N_{max} \,\approx\, \frac{\sqrt{s}}{\Lambda} \,\approx\, \Big( \,\frac{q^2}{x \Lambda^2}\, \Big)^{1/2}\,.
\end{equation}
This places the limit: 
\begin{equation} 
0 \le \rho \le ( N_{max} - 1) \,.
\end{equation}
 
The structure function $F_2$ will be the sum 

\begin{equation}
F_2(x,q^2)\,=\,  F^{(\rho =0)}_2(x,q^2) \,+\, \sum_{\rho =1}^{N_{max} - 1} F^{\,\rho\,}_2(x,q^2)  \,.
\end{equation}

In the limit $ x \to 0 $ we have 
\begin{eqnarray}
 F^{\,\rho\,}_2(x \to 0\,,q^2) &\approx &
F^{(\rho =0)}_2(x \to 0\,,q^2) \,\frac{1}{x^2}\, 
\Big[ \frac{ (\Delta )_{\rho -1}} {(\rho - 1 )!} \, \Big]^2 \,. 
\end{eqnarray}

\noindent Then
\begin{eqnarray}\sum_{\rho =1}^{N_{max} - 1} F^{\,\rho\,}_2(x \to 0\,,q^2) &\approx &
F^{(\rho =0)}_2(x \to 0\,,q^2) \,\frac{1}{x^2}\,N_{max}^{(2 \Delta - 1)}\,.
\end{eqnarray}

\noindent Using these results (and eq. (\ref{FEPS})) we find  

\begin{equation}
\label{Final}
F_2 (x,q^2)\,\approx\, 
\pi C_0 \, {\cal Q}^2 \left( \frac{q^2}{\Lambda^2} 
\right)^{1/2} \, x^{-1/2} \,\,.
\end{equation}

The total cross section is related to the structure function $F_2 $ by 

\begin{equation}
\sigma (q^2 , x)  \, =\, 4 \pi^2 \alpha_{EM} \, \frac{ F_2 (x,q^2)}{q^2}.
\end{equation}

\noindent Then our result for small $x$ implies that 

\begin{equation}
\sigma (q^2 , x)   \sim ( q^2 x^\lambda )^{ \gamma_s - 1}\,,
\end{equation}

\noindent with $ \lambda = 1$ and $ \gamma_s = 1/2 $. So we find a scaling law for the total cross section
different from the case $\Delta' = \Delta$ of  eq.  (\ref{PS1}). This scaling is similar to geometric 
scaling.

It is interesting to compare our scaling with other scalings obtained in different kinematical regimes
(at large $gN$).
In ref. \cite{Polchinski:2002jw}, in the regime  $ \exp \,( - \sqrt{gN }) << x << (gN)^{-1/2}\,$
the DIS structure function is

\begin{equation}
F_2  \, = \,  \frac{2 \Delta + 3}{\Delta +2} \, 
 \frac{\pi^2 \,\zeta \vert C_i \vert^2}{ 2 \, (4 \pi \, gN)^{1/2} \, }\,\frac{1}{x}   \,
\left(\frac{\Lambda^2}{q^2}\right)^{\Delta-1} \,  {\cal I}_{ \, 1 , \, 2\Delta+3}\, \nonumber \\
\label{finalf1f2smallx}
\end{equation}

\noindent where $\zeta $ is a dimensionless constant and 

\begin{equation}
{\cal I }_{\, j , \, n } \equiv   \int_0^{\infty} d \omega  \, \omega^n \, K_{j}^2(\omega) \,
= \, 2^{n-2}\frac{\Gamma(\frac{n+1}{2}+j) \,  \Gamma(\frac{n+1}{2}- j ) \, \Gamma^2(\frac{n+1}{2})}{\Gamma ( n+1) } \, \, . 
\end{equation}    

So the total cross section is proportional to
\begin{equation}
\frac{ F_2 (x,q^2)}{q^2} \sim ( q^2 x^{\lambda}\, )^{ \gamma_s - 1 }\,.
\end{equation}

\noindent This corresponds to a scaling with $ \lambda = 1/\Delta $ and $ \gamma_s \,= 1-\Delta$.
For a baryon in QCD $ \Delta = 3$. If this above result could be extended to this case one would find 
$ \lambda = 1/3 $ and $ \gamma_s \,= - 2$. 

A recent article \cite{Hatta:2007he} considered the regime $\, x <  \exp \,( - \sqrt{gN }) \,$ 
with large $gN$ and $N$ large but finite. 
They studied the problems of saturation and unitarity in the scattering amplitudes, 
problems that are related to geometric scaling. In particular they found, in the vicinity of the
saturation line, a scaling law with  $ \lambda = 1 $ and $ \gamma_s \,= - 1$.

\acknowledgements We thank Edmond Iancu for very important comments. 
The authors are partially supported by CLAF, CNPq and FAPERJ.

 \end{document}